# Analysis of periodicity of extinction using the 2012 geological timescale

Adrian L. Melott and Richard K. Bambach

*Abstract.*—Analysis of two independent data sets with increased taxonomic resolution (genera rather than families) using the revised 2012 timescale reveals that an extinction periodicity first detected by Raup and Sepkoski (1984) for only the post-Paleozoic actually runs through the entire Phanerozoic. Although there is not a local peak of extinction every 27 Myr, an excess of the fraction of genus extinction by interval follows a 27-Myr timing interval and differs from a random distribution at the *p* ~ 0.02 level. A 27-Myr periodicity in the spectrum of interval lengths no longer appears, removing the question of a possible artifact arising from it. Using a method originally developed in Bambach (2006) we identify 19 intervals of marked extinction intensity, including mass extinctions, spanning the last 470 Myr (and with another six present in the Cambrian) and find that ten of the 19 lie within ±3 Myr of the maxima in the spacing of the 27-Myr periodicity, which differs from a random distribution at the *p* = 0.004 level. These 19 intervals of marked extinction intensity also preferentially occur during decreasing diversity phases of a well-known 62-Myr periodicity in diversity (16 of 19, *p* = 0.002). Both periodicities appear to enhance the likelihood of increased severity of extinction, but the cause of neither periodicity is known. Variation in the strength of the many suggested causes of extinction coupled to the variation in combined effect of the two different periodicities as they shift in and out of phase is surely one of the reasons that definitive comparative study of the causes of major extinction events is so elusive.

*Adrian L. Melott. Department of Physics and Astronomy, University of Kansas, Lawrence, Kansas 66045, U.S. A. E-mail: melott@ku.edu*

*Richard K. Bambach. Department of Paleobiology, National Museum of Natural History, Smithsonian Institution, P.O. Box 37012, MRC 121, Washington, D.C. 20013-7012, U.S. A. E-mail: richard.bambach@verizon.net*

### Introduction

Raup and Sepkoski (1984) identified a 26-Myr periodicity in the extinction of marine families over the past 250 Myr. That paper led to what became known as the "Nemesis hypothesis," that a dark, distant companion star to the sun might be responsible for generating comet showers causing periodic extinction events (Whitmire and Jackson 1984; Davis et al. 1984). The reliability of the detection of periodicity was criticized in several ways. For example, Hoffman (1985) and Hoffman and Ghiold (1985) argued that in a stochastic distribution of extinction data a peak would be expected every fourth stage on average and, given the average duration of stratigraphic stages, that would be every 26 Myr. This particular criticism was effectively countered by pointing out that the issue was not the average duration between peaks, but rather the extremely regular duration between peaks; the timing didn't average 26 Myr, it *was* 26 Myr (Raup and Sepkoski 1986b; Gilinsky 1986). Stigler and Wagner (1987, 1988) developed more statistically sophisticated arguments that innate characteristics of the geological timescale forced a periodicity, no matter what the data might be. Those arguments may hold for the original Raup and Sepkoski study using the timescale as understood in 1984 (and using family-level taxonomy with extinctions resolved only to the stage level). But Sepkoski went on to create a more refined and richer database of genus temporal ranges (Sepkoski 2002), in many cases to the substage level. Raup and Sepkoski (1986a, 1988) repeated their analysis and argued that the 26-Myr periodicity of extinction was clear at the genus level as well, but acknowledged that it would take more data to resolve the controversy fully.

Analyses for periodicity in extinction using the more accurate timescale published in 2004 (Gradstein et al. 2004) and using the complete genus record (Sepkoski 2002), and studies using the new Paleobiology Database (http://paleodb.org/), which was not begun until 1998 and only recently achieved a level of data permitting full-scale analysis, continue to show a periodicity similar to the one Raup and Sepkoski noted in 1984. Lieberman and Melott (2007: Fig. 5B) found a peak in fractional extinction intensity at 27 Myr ($p$ = 0.02 in the Rohde and Muller [2005] version of the Sepkoski database). Alroy (2008: Fig. S2) showed that the only peak in periodicity of extinction for the Paleobiology Database (PBDB hereafter) reaching the 95% confidence level is at somewhat less than 30 Myr, but he did not discuss this or report the precise value of the periodicity. Melott and Bambach (2010) analyzed both the Sepkoski genus data and the PBDB for periodicity of extinction using the 2004 timescale and found that both display a periodicity in extinction at approximately 27 Myr extending over 500 Myr. The lengthening of the Mesozoic and Cenozoic by about 3%, and the revised dating of the Paleozoic in the revised timescale (Gradstein et al. 2004) explains why the periodicity changed from 26 to 27 Myr (the difference with earlier results is about 2σ as compared with the bandwidth of the spectral peak) and revealed that the periodicity of extinction discovered by Raup and Sepkoski (1984) for the Mesozoic and Cenozoic persisted through the entire Phanerozoic.

The PBDB uses different occurrence data for genera than the Sepkoski data set and also uses different temporal subdivisions than Sepkoski for binning the data. Also, the substage level for temporal subdivision in the Sepkoski data set uses with many more intervals than the original stage resolution. In addition, newer timescales contain new and often different dates than used in 1984. The fact that the 27-Myr periodicity is present in both the PBDB and the Sepkoski data compilations when newer time scales are used suggests that the periodicity actually exists as a phenomenon and is not an artifact of the 1984 timescale as feared by Stigler and Wagner (1987, 1988). However, a residual 27-Myr periodicity persisted in interval lengths in the 2004 timescale (Lieberman and Melott 2007). The Melott and Bambach (2010) analysis also reveals that although all local peaks of extinction (or even all mass extinction events) do not occur within ±3 Myr of the 27-Myr spacing, the periodicity remains regular throughout the Phanerozoic. Because the orbit of a Nemesis-type solar companion at that lengthy period would be unstable and vary, the "Nemesis hypothesis" cannot, in fact, be the cause of the periodicity of extinction (Melott and Bambach 2010 and references therein).

Melott and Bambach (2011b) also demonstrated that mass extinctions cluster preferentially in the decreasing diversity phases of a 62-Myr periodicity in diversity through the Phanerozoic.

Here we use dating assignments from the 2012 geologic timescale (Gradstein et al. 2012) to examine extinction periodicity. This is crucial because shifts in dates of interval boundaries, many of 1 Myr or more, could have weakened or eliminated the periodic signals seen using earlier timescales.. We examine extinction rates as given by Alroy (2008) from the PBDB, but with revised interval end dates. We also use Fourier analysis on a reduction of the Sepkoski data (hereafter called SEP) which interpolates all less well-resolved genus ranges to substage level by proportional assignment based on the distribution of ranges known at the substage level, again with 2012 dates. The Fourier analysis method was performed on SEP data linearly interpolated to 1-Myr intervals, but was checked against Lomb-Scargle (LS) analysis of the non-interpolated data (Scargle 1982, 1989). Both were implemented using AutoSignal 1.7 software. Fourier analysis is the more familiar, classical textbook method, but requires regularly spaced data. It can be applied to irregularly spaced data but only with judicious use of interpolation. The Lomb-Scargle method is less familiar and less flexible, but generally more correct for data with irregular sample spacing. For this reason it is the only one we use on the PBDB data, with its 11-Myr mean sampling interval. Further discussion of these methodological issues is contained in Melott and Bambach (2011a).

We test for any overall periodicity in extinction intensity and also test the spacing of specific intervals of marked extinction intensity taken as discrete events. Because changes in dates in the 2012 timescale are commonly greater than the ~1-Myr bandwidth of the 27-Myr spectral peaks found using the 2004 timescale, preservation of the signal is a nontrivial test of the hypothesis that intervals of marked extinction intensity relate significantly to the 27-Myr periodicity, as is the appearance of the overall periodicity in the SEP and PBDB data with their different combinations of temporal intervals. In examining intervals of marked extinction intensity we find that the new timescale elevates rates in several substages to make 25 such intervals (Fig. 1), rather than the 19 labeled as mass extinctions in Bambach (2006). Of the total of 25 intervals of marked extinction intensity, six of which occur in the Cambrian, 19 occur between 470 Myr ago and the present, the time span for which we can reliably evaluate overall periodicity patterns. The temporal distribution of these 19 intervals strongly favors both the 27-Myr periodicity and the decreasing diversity phases of the 62-Myr periodicity of diversity fluctuation. Although this study establishes the relationship of marked intensity of extinction to these two periodicities, the causes of the periodicities are still unknown. Several geological features have periodicities similar to the well-known 62-Myr periodicity (Melott and Bambach 2011b; Meyers and Peters 2011; Melott et al. 2012; Prokoph et al. 2013; Rampino and Prokoph 2013), but no consistent geologic features have yet been shown to follow the 27-Myr periodicity.

This paper brings information on the 27-Myr periodicity and the 62-Myr periodicity published elsewhere into the paleontological literature for the first time. It further defines and documents the timing of intervals of marked extinction intensity to both periodicities and illustrates how the interaction of the periodicities as they come in and out of phase may influence the severity of extinction events, thus making systematic study of causes difficult.

**Spectral Analysis of Extinction Rates**

*Methods for Spectral Analysis.*—The SEP data set is one of the data sets used to examine overall biodiversity trends in Melott and Bambach (2011a). It is a complete compilation of all marine genera compiled by Sepkoski as of 1996 and it varies from the published set (Sepkoski 2002) by only a few hundred genera added later out of over 35,000. Bambach interpolated all the less well-resolved data to the substage level, such that the proportions matched those of well-resolved data in each substage bin. With 164 intervals over the Phanerozoic, the temporal resolution is about 3 Myr. Substage boundaries were determined from the 2012 geologic timescale (Gradstein et al. 2012).

The PBDB data are values of the parameter μ, an extinction rate described by Alroy (2008), and provided by him. The PBDB data are grouped into 48 temporal intervals averaging about 11 Myr in duration. The only alteration in the data was using the 2012 geologic timescale for the PBDB bin boundaries.

We plotted the proportion of extinction per substage, using the interval end dates for spacing. Both data sets were truncated at 475 Ma, and detrended by fitting a straight line by least squares regression, and a set of residuals around the baseline was calculated by subtracting the expected values from the data. Because the evolutionary dynamics in the Cambrian and very early Ordovician are so different from those after the Early Ordovician (Bambach et al. 2004; Lieberman and Melott 2007), we omitted the Cambrian and Early Ordovician from the analysis to prevent unreliable detrending and analytical comparison of the time series. Moreover, because the evolutionary dynamics in the Cambrian were unusually volatile, primary analysis only on the post-Early Ordovician portion of the Phanerozoic is reasonable. Trilobites dominated the Cambrian fauna, constituting 50% to 70% of the total number of genera in each time interval, whereas no class-level taxon constituted more than about 15% of total diversity after the Early Ordovician (tabulations of Sepkoski genus data held by Bambach). It is not clear whether a fauna dominated by just one class, especially with the volatility recorded in the Cambrian, would have responded to environmental perturbations in a manner comparable to the more evenly structured multi-taxon faunas with the less volatile evolutionary dynamics of later times. Similarly, Wang and Bush (2008) omitted several classes confined to the Cambrian from their study on adjusting global extinction rates simply because generalizing on extinction rates for them is not practical.

The time-series analysis examines variations around the broad trend line of reduction in the rate of extinction, a feature first emphasized by Raup and Sepkoski (1982). The variations were examined in two ways. We use the Fast Fourier Transform (FFT) method for the SEP data, which has time resolution amply adequate for the detection of a 27-Myr period. Because FFT requires evenly spaced data, we must interpolate between the variously spaced endpoints of the substages used by SEP. Because the SEP intervals average 3.44 Myr in duration, interpolation is acceptable. We are not examining periods comparable to or smaller than twice the mean interval length, called the Nyquist interval. Therefore, the SEP data were linearly interpolated between adjacent data points in order to create a regularly spaced sampling amenable to Fourier analysis. A second, related analysis by Lomb-Scargle (which does not require interpolation or evenly spaced data) also was performed on the SEP data prior to interpolation. However, at the present AutoSignal implementation of LS does not provide significance levels. We consider only the Lomb-Scargle method appropriate for the PBDB data, as it is severely limited in time resolution with its ~11-Myr mean interval length. Description, discussion, and testing of both procedures are contained in various textbooks, as well as our previous paper (Melott and Bambach 2011a) and references therein.

*Spectral Peaks in Individual Data Sets.*—Figure 2 shows plots of spectral power versus frequency (in units of per Myr) on a linear scale, derived from the given analysis procedure. Both data sets show a peak at 27 Myr using LS (Fig. 2A,C) as FFT does also for SEP (Fig. 2B). Although not as statistically robust, PBDB does for FFT as well (not shown). We note that these results are valid only for frequencies up to the Nyquist frequency, equivalent to the inverse of twice the sampling interval. This is about 0.17 $Myr^{-1}$ for SEP and 0.048 $Myr^{-1}$ for PBDB, corresponding to 6 Myr and 21 Myr, respectively. The signal we are testing is at 0.037 per Myr, so the PBDB choice of intervals strains the time resolution, but does fall within the range for valid analysis. The Lomb-Scargle results shown here are truncated at the appropriate place.

The series of curved lines in Figure 2B indicate confidence levels of 95%, 99%, and 99.9% against the spectral peak arising from randomly distributed data. All figures were obtained using AutoSignal 1.7 software, and the 27-Myr peak in Figure 2B rises to about 98% confidence. It is important to note what these results do and do not imply. First, they are measurements of the various frequencies that are present in the data, after the removal of the background of overall decline in extinction rate. The results imply the existence, with about 98% confidence inconsistent with noise, of a 27-Myr repeating signal in the extinction rates. The signal is produced by the distribution in time of all rates of extinction, both large and small, not just those regarded as mass extinction events: for example small peaks at about 227 Ma and 307 Ma among others.

Power spectra do not show phase information. The timing of the peak components from the two data sets is different by about 4 Myr at the Holocene, but because their frequencies are slightly different, their timings nearly coincide at the time of the end-Permian extinction. No doubt this very large perturbation "anchors" the two series. Because of the coarser set of time intervals involved in the PBDB data, as well as the fact that the dates for PBDB last appearances are the last occurrences entered from a selection of general literature—rather than the last known occurrence for each genus (which is what the Sepkoski data records)—the small phase difference for the two data sets is not a concern.

The most important and durable objection to the original Raup and Sepkoski publication was raised by Stigler and Wagner (1987, 1988), who suggested that the periodicity was implicit to the timings of interval boundaries in the geological timescale then in use. This is not a fatal objection; boundaries between intervals are typically placed when there are many faunal changes, often coincident with extinctions. Most major stratigraphic boundaries were identified before radiometric dating was developed and absolute dates could be attached to them. As noted above, several studies (Lieberman and Melott 2007: Fig. 5B; Alroy 2008: Fig. S2B; Melott and Bambach 2010) using the 2004 version of the timescale identified an extinction intensity peak at 27 Myr, and Lieberman and Melott also noted and discussed a weak peak in the power spectrum of stratigraphic intervals at 27 Myr. Because the interval lengths have changed in the 2012 geologic timescale, our present work constitutes a further test of the idea that stratigraphic intervals are intimately tied to the appearance of this extinction periodicity, as well as a test of whether they are, in general, robust against the shifting of interval boundaries. We have examined the power spectrum of the length of the intervals used in our analyses (Fig. 3). We find that the weak peak at 27 Myr found by Lieberman and Melott (2007) is now greatly reduced, and is below neighboring spectral peaks; no peak appears at all near 27 Myr in the spacing of the PBDB intervals used here (see arrows in Fig. 3). Thus, there is no longer any hint of a connection of interval spacing to any 27-Myr extinction periodicity we may find.

In summary, the power spectrum for interval length using dates determined with the 2012 timescale shows no connection of interval spacing to a 27-Myr extinction periodicity, and both SEP and the PBDB show a 27-Myr periodicity in extinction from essentially independent data compilations. Although this periodicity is not nearly as strong as the 62-Myr periodicity in diversity fluctuation documented by Rohde and Muller (2005) and further explored by Melott and Bambach (2011a,b and several earlier studies by Melott and colleagues referenced therein), it does rise to the level of moderate statistical significance ($p$ = 0.02) and merits examination.

*Check against Additional Data Reductions*—We examined two additional compilations as a cross-check. The first are the two reductions of the SEP data by Shanan Peters, available at http://strata.geology.wisc.edu/jack/ that we use below. These are reductions of the data published in Sepkoski (2002) but include only the genera in which both endpoints (first and last) are known at the substage level. They differ from Rohde and Muller's "well-resolved" data (also used below, and for which a 27-Myr periodicity is known [Lieberman and Melott 2007]) in that Rohde and Muller interpolated endpoints that were resolved at stage level to the substage. Peters's sorting routine does not interpolate and excludes any genus that does not have both endpoints resolved at the substage level. Therefore those data sets are smaller than the others we analyze. The two SEP reductions from Peters's website (with and without singletons) show the periodicity, but the 27-Myr peak is not significant because the numbers are inconsistent with the total number of extinctions that did occur. For instance, if no extinctions in a stage are resolved to substage level, no extinctions are noted for any substage in the stage; but when unresolved data are interpolated into the sets of substages, they represent extinctions that actually did occur at some time during those substages. In effect, accuracy is sacrificed for precision in the two reductions of SEP data from Peters.

The second cross-check set is the data from Wang and Bush (2008) that we also mention below. In the Wang and Bush (2008) data, the 27-Myr peak rises in significance to a level $p$ << 0.001 in both their observed and adjusted data. Wang and Bush used a more condensed set of time intervals than the complete set of substages, and the combination of some substages into longer intervals, especially those containing major extinction events, enlarged the differences between low and high extinction intervals; moreover, because a significant number of major extinction events do follow the 27-Myr spacing, the reduction in number of intervals clusters extinctions in such a way that the periodicity appears enhanced. At any rate, the main conclusion from the examination of these two reductions is that the 27-Myr extinction periodicity appears at some level in each.

### Faults with a New Objection to the Statistical Inference of Periodicity

As this manuscript was being prepared, a paper appeared (Feng and Bailer-Jones 2013, hereafter FB-J) that discussed the fit of biodiversity data to galactic motion and simple periodicities. The authors argue that, when properly treated, statistical analysis does not give any support for periodicity or coincidence with astronomical, particularly galactic, motion of Earth. There appear to be a number of problems with this study. We discuss these issues more completely in Melott and Bambach (2013), but briefly summarize three of the paleontologically relevant ones here.

1. FB-J discuss the 62-Myr periodicity in marine biodiversity first uncovered by Rohde and Muller (2005). However, in the entire paper they only examine extinction rates and events, whereas this periodicity results from the coherent interaction of both extinction and origination fluctuations, producing a stronger signal than either would or could alone (Melott and Bambach 2011b).

2. FB-J examine the fractional extinction rate, inferred from the Sepkoski data as reported by Bambach (2006), and inferred from the PBDB data as reported by Alroy (2008). They also examine the timing of mass extinctions also as reported by Bambach (2006). In doing so, there is no indication in the text that they have utilized the 2012 geological timescale.

3. More seriously, they have assigned both mass extinction events and "continuous" extinction rates to the middle of the substage in question. Interestingly, Figure 4 demonstrates that there is almost no relationship between interval length and magnitude of extinction during an interval ($r^2$ = 0.0003). Although it has been argued before that extinction occurs predominantly in pulses at times of faunal turnover (Vrba 1985, 1993; Morris et al. 1995), many of the pulses documented in those papers occur within the substage and stage intervals we are dealing with (intervals in the Sepkoski data average 3.44 Myr in length; those in the PBDB average 10.86 Myr in length). But with virtually no connection between total extinction in an interval and interval length, neither continuous extinction nor frequent small pulses of extinction govern the differences in total extinction among time intervals. However, most interval boundaries are placed at times of marked faunal change, aiding in correlation, and we know that much extinction is concentrated at the ends of intervals, unambiguously so for well-known major mass extinctions and apparently for most other intervals as well (Foote 2005). Because intervals range from 7 to 18 Myr for PBDB, and from 2 to 10.8 Myr for the SEP data, and because much of the difference in extinction magnitude among intervals is due to pulses of extinction at interval boundaries, the assignment of extinction values to interval midpoints results in a systematic and varying error of a shift to earlier dates, which would be expected to degrade the signal.

For these and other reasons, we do not accept the conclusions of FB-J. In Melott and Bambach (2013: Fig. 2) we show the power spectrum of SEP extinctions using two different treatments of the data: when the 2004 geological timescale is used with data assigned to the midpoints of intervals, rather than using the 2012 timescale and assigning dates to ends of intervals, amplitude is degraded, and the significance drops from $p$ = 0.02 to $p$ = 0.2. Furthermore, in Melott and Bambach (2013) we conducted an assessment using Akaike weights and showed that in spite of these problems, the maximum likelihood data presented by FB-J still give an approximate 90% probability that a simple periodic model for mass extinctions is preferred over any of 18 others they consider.

**The Timing of Mass Extinctions and Other Intervals of Marked Extinction Intensity**

*Background Considerations.*—For the 27-Myr periodicity of extinction there is some uncertainty related both to the width of the peaks in our spectral analysis (which are quite narrow, but do have finite width; see Fig. 1) and to the dating of stratigraphic boundaries and extinction events in the geologic record. The latter uncertainty involves both experimental error in radiometric dating and potential estimation error because most ages for interval boundaries must be interpolated from the precisely dated tie points in the timescale, even if for only small intervals of time (Gradstein et al. 2012). Therefore, we use a ±3-Myr "window" around the exact 27.16-Myr spacing of the periodicity of extinction, with the most recent peak at 9.5 Ma, as the age range we consider associated with the maximum expected extinction. Any event coming within 3 Myr of a peak in the periodicity is considered a "hit." Because the "window" occupies six of the 27 Myr of each cycle we would expect 0.222… of events to lie within such a window if they were randomly distributed. Likewise, a second temporal zone of potential interest is the minimum midway between each 27-Myr maximum in the periodicity. We need to see how intervals of high extinction relate to that low point in extinction as well and refer to the ±3-Myr interval centered on the midpoints 13.58 Myr from each 27-Myr peak as the "gap."

For the 62-Myr periodicity the portions of interest are the 31-Myr segments from the diversity maxima to diversity minima, and the expected likelihood of random distribution of intervals in these half-cycles is 0.5. Because diversity increases when origination exceeds extinction and decreases when extinction exceeds origination we would expect intervals of higher extinction to be concentrated in the decreasing diversity phases; as noted above, Melott and Bambach (2011b) observed that mass extinctions cluster in phases of declining diversity in the 62-Myr periodicity. The latest peak in the 62-Myr periodicity occurred 27.8 Myr ago.

We also specify intervals and events that we consider of interest for their association with high amounts of extinction. Paleontologists studying faunal turnover have recognized varying numbers of extinction events. Barnes et al. (1996) list some 62, spanning from the current human-associated ecological crisis back to the Precambrian/Cambrian boundary event. At the other extreme, Bambach et al. (2004) were able to statistically justify recognizing only three of these extinction events as unusual by elevated magnitude of extinction alone (the end-Ordovician, end-Permian, and end-Cretaceous). When diversity loss is considered, five events (the end-Ordovician, the end-Frasnian in the Late Devonian, the end-Permian, the end-Triassic, and the end-Cretaceous) qualify as unusual intervals of diversity depletion (Bambach et al. 2004). These were the only times when more than 20% of the diversity (measured as number of genera) present at the beginning of an interval was lost during the interval. No other intervals lost more than 13.2% of their diversity, so these five diversity depletions really are "the big five" as initially recognized by Raup and Sepkoski (1982).

For tallying intervals of marked extinction intensity the 62 events listed in Barnes et al. (1996) are not selective enough. Many of those events are minor—some are restricted to just one or two groups, some are only regional, and some are small-magnitude events that cannot be regarded as times of severe general extinction risk (20 of the events listed in Barnes et al. 1996 do not show as peaks of extinction for the substage in which they occur in any of the four synoptic compilations of extinction data used below). On the other hand, three or even five unambiguously large events are not sufficient to do a robust statistical evaluation.

In a review paper on mass extinctions Bambach (2006) compared local peaks of the magnitude and rate of extinction in four differently selected sets of extinction data derived from the Sepkoski genus database and found that the four sets of data only shared peaks of both magnitude and rate in 19 of 164 substage intervals. Because those 19 intervals had local peaks for both measures of extinction intensity in all four compilations (each of which was constructed somewhat differently) Bambach used the universality of locally higher extinction intensity as a criterion for designating those intervals as times of mass extinction. We have chosen to revisit that method of evaluation for two reasons: (a) it needs reexamination because the new timescale affects the rate calculation, and (b) it provides a way of designating intervals of interest without making individually subjective decisions.

*Designating Intervals of Marked Extinction Intensity.*—A unique advantage of the Sepkoski data set is the availability of four different compilations based on the Sepkoski data, each using a different method of selecting data, which allows us to differentiate intervals of marked extinction intensity from random variation in the data.

The four compilations of extinction data are as follows:

1. The data set identified above as SEP. It comprises all genera in the Sepkoski data set including "singletons" (genera reported from just a single stratigraphic interval) and proportionally assigns those range endings not recorded at the substage level to appropriate substages on the basis of the proportion of range endings known to occur in the substages included in the larger interval specified by the poorly resolved range ending. This method increases the number of genera ending in each time interval above those known by completely resolved range endings but respects the pattern of well-resolved data. The data are distributed in 164 substages.

2. The data compiled by Rohde and Muller (2005). It uses only genera with range endings resolved to either substage or stage level and omits singletons and genera with range endings less well resolved than to stage. Stage-level range endings are divided evenly, not proportionally, among the substages of the stage; i. e., if there are two substages, a stage-level range ending is assigned, half to the first and half to the second of the two substages; with three substages, a stage-level range ending is divided into thirds, and one-third is assigned to each substage. There are 165 substages in the Rohde and Muller compilation (they resolved three intervals in the Cenozoic into substages that Sepkoski did not, and combined two Cambrian substages that Sepkoski used into one stage).

3. Data obtained from the website for the Sepkoski database maintained by Peters (http://strata.geology.wisc.edu/jack/) for only those genera resolved at both the start and end of their range to substage (thus omitting all with either first or last part of range not given at the substage level) and with singletons included. These data were described in more detail earlier. There are 156 substages listed in Peters's compilation (Peters combines several substages that Sepkoski used).

4. Data obtained from the website for the Sepkoski database maintained by Peters as in (3) but omitting singletons; thus all genera occurred in multiple substages. This is the most restricted set of data.

Using these four compilations allows us to compare extinction data for 156 intervals of time, ranging from using all the data available (but interpolating a fairly large number of range endings) to using only a very selective grouping of genera that ranged through several substages and have both ends of the range well known. We used two metrics : (1) magnitude of extinction, expressed as the proportion of extinction (the number of genera with ranges ending in an interval divided by the total number of genera occurring in the interval), and (2) rate of extinction, expressed as the proportion of extinction per million years (proportion of extinction in an interval divided by the duration of the interval in millions of years). The reason rate is used as well as magnitude is to prevent long intervals, during which simple "background" extinction can accumulate considerable total extinction, from being grouped with intervals in which stresses caused more extinction than would normally occur over that interval of time (see Fig. 3). However, because extinction is commonly pulsed and often concentrated at or near the end of intervals (Foote 2005), we considered the timing of extinction in an interval to be at the end of the interval, unless a specific date for an extinction event has been independently determined. An example of the latter is dating the Toarcian extinction event in the earliest Toarcian, at about 182 Ma, rather than at the end of that substage, 178.4 Ma (and, in this instance, the high extinction values in the Pliensbachian are now also understood to be exacerbated by "Signor-Lipps Effect" of "smear-back" from the extinction event only a few hundred thousand years into the early Toarcian).

With these four different data sets and 2012 temporal data we used the criterion that a local peak of both magnitude and rate of extinction in all four data sets is required to designate an interval as being of marked extinction intensity. Thus these intervals can be designated independently, without regard to previous recognition as extinction events or mass extinctions. Twenty-five intervals qualify, with six in the Cambrian and 19 occurring in the last 470 Myr (Table 1, Fig. 5).

Although Bambach (2006), using the 2004 timescale, found that each "qualifying" interval contained a major extinction event that could justify the title of mass extinction, the same is not true for the new evaluation using the 2012 timescale. All 19 of the mass extinction intervals of 2006 do persist as intervals of marked extinction intensity, but six additional intervals also have local peaks of both magnitude and rate of extinction in each of the four compilations, two in the Cambrian (the end-Tommotian and end-Tojonian) and four occur in the last 470 Myr.

The six intervals (Table 1) not previously singled out as major extinction events are (1) the end-Tommotian and (2) the end-Tojonian in the Cambrian; (3) the middle Caradocian (a stage name no longer in current use for global correlation, but one used by Sepkoski) in the Ordovician; (4) the late Moscovian in the Late Carboniferous; (5) the Spathian in the Early Triassic; and (6) the middle Miocene (the combined Serravalian and Langhian stages) in the Cenozoic. Some record of extinction or faunal instability is associated with each. (1) and (2) are both noted as extinction events in Gradstein et al. (2012: p. 470 and references therein. (3) Patzkowsky et al. (1997) discuss an extinction in North America that accompanied a carbon isotope excursion in the Middle Caradocian and note that this event appears more like the Cambrian biomeres than the massive end-Ordovician mass extinction and Ainsaar it al. (2004) report similar faunal turnover and environmental disruption in Baltoscandia. (4) Sahney, Benton and Falcon-Lang (2010) relate major reorganization of terrestrial faunas to the collapse of Late Carboniferous rainforests and major climatic shifts at the end of the Moscovian, and Stanley and Powell (2003) show a definite small extinction peak for the marine fauna in the Moscovian. (5) The Spathian is an interval late in the Early Triassic that begins with a major carbon isotopic anomaly, the last major perturbation during the unstable time following the extraordinary end-Permian mass extinction, and on land the flora changed as climate changed during the interval (Galfetti et al. 2007). Marine faunas were also undergoing considerable change in the Spathian as the depauperate survivor fauna was replaced by the first diverse fauna of the Mesozoic (Hoffman et al. 2013). (6) In Sepkoski's data the middle Miocene has always appeared as an interval of high extinction relative to most of the Cenozoic, but it has seldom been publicized as a time of unusual extinction intensity. However, recent work on detailed climatic modeling demonstrates that the Middle Miocene climate fluctuated considerably, attaining an "optimum" of higher temperatures than had occurred for a considerable time previously, and this optimum ended rather abruptly as cooling leading eventually to the Pleistocene ice ages began (Böhme 2003). Unstable oceanic chemical conditions are also indicated by widespread deep-sea unconformities during the mid-Miocene (Ujiié 1984). All six events reasonably qualify as additional intervals of marked extinction intensity because they each display a local peak of both magnitude and rate of extinction no matter how the Sepkoski compilation is parsed, just as is the case for known intervals containing mass extinctions.

We use the criterion of a local peak in both magnitude and rate in each of the four data compilations to avoid including too many random peaks, because random peaks would be expected in any series of 150+ data. Each individual data set had between 37 and 40 intervals with local peaks for both magnitude and rate of extinction, yet only 25 intervals have local peaks for both in each data set. This set of substage intervals incorporates all the intervals that would appear to have marked extinction intensity no matter how the data are selected. Interestingly, *only* the set of intervals sharing peaks of both magnitude and rate of extinction in each data set show a clear preference for the two periodicities of extinction.

   *Timing Patterns for Intervals of Marked Extinction Intensity.*—Because of the peculiar high evolutionary volatility of the fauna in the Cambrian (e.g., Lieberman and Melott 2013), we will examine the timing of extinction events primarily for just the last 470 Myr, as noted in the spectral analysis section of this paper, although some appropriate observations on the Cambrian will be mentioned in the discussion section below. Sixteen of the 19 intervals of marked extinction intensity in the last 470 Myr fall within phases of decreasing diversity associated with the 62-Myr periodicity and only three occur during phases of increasing diversity (Table 1, Fig. 5). This is a highly unlikely balance, with a binomial $p$ = 0.002. Ten of the 19 intervals of marked extinction intensity occur within ±3 Myr of the 27.16-Myr spacing, too. To have over half the occurrences in just 0.222 of the available time is also highly unlikely, with $p$ = 0.004. Along with that, none of the intervals of marked extinction intensity occur in the ±3-Myr "gap" around the spacing point halfway between the 27.16-Myr spacing points ($p$ = 0.01).

   Of the 19 mass extinctions designated in Bambach (2006), 15 occur during the 470 Myr considered here. All 15 are included in the intervals of marked extinction intensity, but four of the 19 intervals of marked extinction intensity are not designated mass extinctions. However, these four intervals all occur in the decreasing diversity phases of the 62-Myr periodicity and two of the four occur within ±3 Myr of the 27-Myr spacing. All three intervals of marked extinction intensity that occur during increasing diversity phases of the 62-Myr periodicity are designated mass extinction events; two of the three are within the ±3-Myr "window" around the 27-Myr spacing of extinction and the third is "close" at 3.7 Myr (Fig. 5). Of the more prominent mass extinctions, all three of the most severe (actually four points, because the end-Ordovician is two separate events) occurred during both the decreasing diversity phases of the 62-Myr periodicity and the ±3-Myr window around the 27-Myr extinction periodicity spacing points. All five of the "big five" mass depletions of diversity are in decreasing diversity phases of the 62-Myr periodicity, and four of the five are in the ±3-Myr window around the 27-Myr spacing points. The one that is not is the end-Frasnian extinction in the Late Devonian, which is at a distance of 9.62 Myr and, interestingly, is the diversity depletion event with the greatest dependency on origination failure and the least dependency on elevated extinction of the "big five" events (Bambach et al. 2004; Rode and Lieberman 2004; Stigall 2012).

   The clear relationship of intervals of marked extinction intensity, including mass extinctions, to the 62-Myr and 27-Myr periodicities does not hold for the extensive listing of extinction events in Barnes et al. (1996). For that list, in which 55 of the 62 are in the last 470 Myr, 36 events do not occur during intervals of marked extinction intensity if our criteria are used. Of those, 20 occurred in increasing diversity phases and 16 in decreasing diversity phases of the 62-Myr periodicity, close to an even split, and the 36 events show no strong relationship to either the 27-Myr spacing points or the midpoints between them. The bottom line is that intervals of marked extinction intensity, including mass extinctions, (a) preferentially occur near the 27-Myr spacing points, (b) are rare near the midpoints of that spacing, and (c) are strongly associated with decreasing diversity phases of the 62-Myr periodicity, but extinction events that are smaller or more taxonomically restricted are not.

   *Timing Relationships for Intervals with Peaks of Extinction in Some, but Not All, Data Compilations.*—With the very strong association of intervals of marked extinction intensity both with the decreasing diversity phases of the 62-Myr periodicity and with the 27-Myr periodicity, one would expect that intervals that "almost" qualify as intervals of marked extinction intensity might also show some preference for the same relationship. If having eight local peaks (four for magnitude and four for rate, using the four different sets of data) connects those intervals to the periodicities of extinction, shouldn't intervals with seven, six, or five local peaks (a majority of possible local peaks in the four data sets, thus requiring at least one data set to have peaks of both magnitude and rate) also have somewhat similar behavior to intervals in which all four data sets record local peaks of both magnitude and rate? There are 14 intervals in the last 470 Myr that have five to seven local peaks of magnitude and/or rate of extinction in the four data sets examined. Only one is within ±3 Myr of the 27-Myr spacing points and three are in the "gap" range, thus showing no preference whatsoever in regard to the 27-Myr spacing. On the other hand, 11 of the 14 occur during increasing diversity phases of the 62-Myr periodicity (for a $p$ = 0.029) and just three occur in the decreasing diversity phases, the opposite relationship to that for  intervals of marked extinction intensity. This suggests that conditions (as yet unknown) during decreasing diversity phases of the 62-Myr periodicity elevate extinction intensity enough to "recruit" what otherwise might have been five to seven peak intervals into the cadre of intervals of marked extinction intensity. Severe "insults" can occur at any time (the pulse aspect of extinction events as envisioned by Arens and West [2008] and Feulner [2011]), but the "press" aspect that might increase general likelihood of extinction may be more general or intense during the decreasing diversity phases of the 62-Myr periodicity, thus creating the imbalance in distribution of intervals of marked extinction intensity, leaving an excess of "almost qualified" intervals in the increasing diversity phases.

The 35 intervals with one to four local peaks scattered among the four data sets show no positive preference for either periodicity. Nonetheless, although about the number expected in a random association occur in the ±3-Myr window around the 27-Myr spacing points, 13 of 35 occur in the ±3-Myr window around the midpoint between 27-Myr spacing points, a significant concentration ($p$ = 0.033). However, because six of the ten intervals with just a single local peak in any of the four data sets occur in the ±3-Myr "gap" midway between the 27-Myr spacing points ($p$ = 0.01) it is simply the addition of those six occurrences to the seven (of 25 [ $p$ = 0.31]) of the two-to-four-peak set that creates the imbalance in the "gap" portion of the 27-Myr pattern. Apparently the 27-Myr system also promotes extinction intensity, as does the 62-Myr periodicity, because the 10 single peak intervals are evenly split between the increasing and decreasing phases of the 62-Myr periodicity, but are significantly more common in the "gap" windows (and equivalently less common in the rest of the 27-Myr cycles) than expected. A stress system outside the gap region associated with the 27-Myr periodicity apparently "promoted" what would have been single-peak intervals to intervals with multiple peaks, leading to the remnant excess of single-peak intervals in the "gap" spacing.

The 67 intervals with no peaks in any of the four data sets examined are randomly distributed through time and show no preferential relationship to either periodicity.

*Summarizing the Influence of the Two Periodicities.*—The preferential occurrence of intervals in relation to the two periodicities seems to be entirely governed by the effects that promote extinction beyond the level from "pulse" causes, and these conditions occur primarily during the decreasing phases of diversity in the 62-Myr periodicity and during the ±3-Myr window around the 27-Myr spacing points in the 27-Myr periodicity, with the least expected extinction in the increasing diversity phases of the 62-Myr periodicity and in the "gap" midway between the 27-Myr spacing points. The frequency of occurrence of intervals of marked extinction intensity in phases of decreasing diversity in the 62-Myr periodicity is very strong (Table 1), with a likelihood of only $p$ = 0.01 for the last 470 Myr and $p$ = 0.0005 if one includes the Cambrian extinction events as well. However, three of the intervals of marked extinction intensity do occur during phases of increasing diversity (four if the Cambrian is included). For the 27-Myr periodicity, ten of 19 intervals falling within the ±3-Myr window may not seem as clean as the 16 to 3 ratio in the 62-Myr periodicity, but the occurrence of over half of the intervals in just 22.2% of the time span of the periodicity also produces a $p$ = 0.009, equivalent to that for the 62-Myr periodicity. Both periodicities enhance extinction, but don't force universal response. Neither dominates over the other. They are most effective when in phase. Both are pervasive, but both are apparently subtle "press" drivers (in the sense of Arens and West's press-pulse theory of extinction), not severe "pulse" events.

## Discussion

*Intervals of Marked Extinction Intensity in the Cambrian Period.*—Because of issues related both to correlation problems between the currently used stage designations and Sepkoski's old choices of stratigraphic intervals and to the exceptionally volatile evolutionary dynamics of the Cambrian and Early Ordovician, we ran our analysis of periodicity of extinction for just the past 470 Myr. However, one might expect such periodicities to continue back to earlier times. What intervals of higher extinction stand out in the Cambrian and how do they relate to the two periodicities?

Because the Cambrian fauna had simpler ecosystem structure (Bambach et al. 2007) as well as overwhelming diversity dominance by just one class (the trilobites) the Cambrian may not have extinction patterns exactly comparable to the rest of the Phanerozoic (e.g., Lieberman and Melott 2013). Nonetheless, six extinction events in the Cambrian can be linked to substage data in the Sepkoski data set (qualifying them as intervals of marked extinction) and each is specifically noted in the charts in the Cambrian chapter in Gradstein et al. (2012); thus the dates of these six events can be located in relation to the two periodicities despite uncertainties about dating their Sepkoski interval boundaries (Table 1). No other time period had more than three such intervals, another reason for suspecting that unique Cambrian dynamics are involved.

Three of these events fall in the "gap" interval of ±3 Myr around the midpoint between 27-Myr spacing points, which is not the case for any of the 19 intervals of marked extinction intensity from 470 Ma to now, and one of these even occurs in an increasing diversity phase of the 62-Myr periodicity. However, five of these Cambrian events occur within a decreasing diversity phase of the 62-Myr periodicity and one event (the Marjuman/Steptoan biomere extinction) does lie within ±3 Myr of a 27-Myr spacing point. Although these results appear to weaken the relationship between the 27-Myr periodicity of extinction and the temporal positioning of intervals of marked extinction intensity, the probability of having 21 out of 25 such intervals fall into the decreasing diversity phases of the 62-Myr periodicity increases to a highly significant $p$ = 0.0005. The conditions that encourage diversity loss during these phases may have been sufficient to promote some intervals in the Cambrian, with its uniquely high evolutionary volatility, to the level of marked extinction intensity even though they were not also close to the preferential timing of ±3 Myr around the 27-Myr spacing.

*The Decreasing Diversity Phase in the Devonian.*—Over the Phanerozoic there are nine phases of decreasing diversity related to the 62-Myr periodicity, but only one has no interval of marked extinction intensity that also lies within ±3 Myr of a 27-Myr spacing point. That phase is in the Devonian (Fig. 5, at 399 to 368 Ma). Even though much of the Devonian took place during a phase of decreasing diversity in the 62-Myr periodicity, the Early to Middle Devonian marks the time of highest diversity during the entire Paleozoic (Bambach 1999; Bambach et al. 2004; Alroy 2010).

The subtle nature of the influence of the two periodicities on diversity and evolutionary dynamics may be illustrated by their apparently small influence on the Devonian fauna. The Devonian is well known as a time of considerable faunal turnover. Bambach (1999) emphasized that the peak in diversity in the Devonian was actually the overlap of two suites of dominant taxa: one had been important since the Ordovician and the second became dominant after the multiple extinction events in the Middle and Late Devonian. A major radiation in vegetation, including tree-size plants becoming dominant on land for the first time, also altered the carbon cycle and the supply of nutrients from the land to the nearshore oceans as the turnover in the marine fauna was also taking place. The influences of the two periodicities described in this paper may have been obscured by these complex biological interactions. Those interactions may have been just as important as—or even more important than—physical influences in producing both the turnover of dominant taxa and extinction during the Devonian, possibly including the Frasnian/Famennian event, the only one of the "big five" diversity depletions that does not occur within ±3 Myr of one of the 27-Myr spacing points (Table 1).

*The Complex "Causal" Question.*—What causes major extinction events and/or intervals of marked extinction intensity? The variety of environmental disruptions hypothesized to be possible causes of major extinction events include (a) ocean acidification (Hönisch et al. 2012; Kiessling and Simpson 2010; Pelejero et al. 2010; Pörtner 2008), a popular issue because of its relevance to modern environmental issues; (b) oceanic euxinia (Meyer and Kump 2008), an extreme end-member of conditions leading to acidification and anoxia; (c) volcanism, especially the eruption of large igneous provinces and the emission of noxious volatiles (Sobolev et al. 2011); (d) asteroid and/or comet impacts (White and Saunders 2005; Glikson 2005), including possible association with flood basalt volcanism; (e) both warming and cooling events (Twitchett 2006; Mayhew et al. 2008); (f) sea-level change (Hallam and Wignall 1999), an older idea as a potential cause of extinction, but one that can both eliminate habitat and bring anoxic or euxinic water into shelf environments; and (g) effects of astrophysical ionizing radiation on the atmosphere and biota (Melott and Thomas 2011). All these reviews of possible causes show some interesting correlations, but none can account for all major events and few actually demonstrate an unambiguous cause-effect relationship for specific events. These reviews also demonstrate that there has been little success relating the magnitude of disturbances to the severity of associated extinctions. Wang and Bush (2008) demonstrated convincingly that evolutionary turnover has a direct influence on the raw rate of extinction, with more evolutionarily volatile taxa being replaced by less volatile taxa as extinction rates have declined. To compare the actual severity of extinctions at different times one would need to correct for the sensitivity of the fauna at each time. The same severity of event could cause different severity of extinction, depending on the intrinsic sensitivity of the fauna of the time to the stress imposed.

The two periodicities that influence extinction add to the puzzle. Two examples related to climate change illustrate how the difficulties in relating potential causal phenomena with severity of extinction are exacerbated by the influence of the periodicities considered here.

1. The onset of large-scale glaciation correlates with several extinction events of different magnitude, from the enormous end-Ordovician pair of events (the pair together constitute one of the "big five") (Finnegan et al. 2012) through the significant, but not spectacular, mid-Carboniferous (Serpukhovian) event (McGhee et al. 2012) to the smaller end-Eocene extinction event (Houben et al. 2012). Why should these three events, which are similar in climatic influence, have been so different in biotic effect? One possibility is that the two end-Ordovician events occurred during a decreasing diversity phase of the 62-Myr periodicity as well as in the window within ±3 Myr of a 27-Myr spacing point, whereas the Serpukhovian extinction occurred during a decreasing diversity phase of the 62-Myr periodicity but, at 8.37 Myr from a 27-Myr spacing point, outside the ±3-Myr window; the end-Eocene, although it fell within ±3 Myr of a 27-Myr spacing point, is one of just four intervals of marked extinction intensity that occurred during an increasing diversity phase of the 62-Myr periodicity. The same general level of physical "insult" may have had different effects based on where in each cycle the insult happened.

2. The same relationship appears in a set of three intervals of global warming: the Paleocene-Eocene Thermal Maximum (PETM), a sudden heating event (McInerney and Wing 2011); the thermal maximum in the Cenomanian–Turonian in the Cretaceous (Huber et al. 2002), and the end-Permian–Early Triassic, in which exceptionally high temperatures are reported (Sun et al. 2012). The PETM has little associated extinction, except for some planktonic foraminifera; the Cenomanian/Turonian boundary is marked by a distinct, but rather small, mass extinction; and the end-Permian extinction was the most devastating in the Phanerozoic. Were the warming events of sufficiently different magnitude to account for the variation in extinction among these three cases or were the relationships of these three events to the two extinction periodicities involved as well? The end-Permian is within ±3 Myr of a 27-Myr spacing point and occurred during a decreasing diversity phase of the 62-Myr periodicity. The Late Cenomanian extinction event also occurred within ±3 Myr of a 27-Myr spacing point but occurred during a phase of increasing, not decreasing, diversity in the 62-Myr periodicity. The PETM occurred outside the ±3-Myr window around the closest 27-Myr spacing point (missing the spacing time by 7.82 Myr) and also occurred almost 2 Myr into an increasing diversity phase of the 62-Myr periodicity. The relationship of the timing of the two periodicities may have influenced the severity of extinction events with similar driving mechanisms.

In the Arens and West (2008) "press-pulse" theory of mass extinction the intensity of extinction is governed by a broadly applied cause of instability for ecosystem structure (the press) and more specific, shorter-term pulses of disturbance generating events that created severe physiological difficulty. Although we agree with Feulner (2011) that periodic effects on extinction are probably relatively modest in strength compared to disruptive environmental events, which are likely to be temporally random in distribution, the fossil record for extinction is clearly influenced by both the 27-Myr and 62-Myr periodicities. By enhancing (or promoting) greater extinction these periodic influences determine when most intervals of marked extinction intensity occur. The various physically and physiologically disruptive events (climate change, eruption of large igneous provinces, events of anoxia, euxinia, ocean acidification, etc.) are "pulse" events, but "press" phenomena associated with the two periodicities appear to have influenced the severity of extinctions triggered by these and other pulse events.

What physical mechanisms are related to periodic causality? At present none are definitively identified, although a variety of both astronomical and geological phenomena offer possibilities. Our previous work, as noted earlier, may rule out the Nemesis model of a dark stellar companion. However, other astronomical models are possible.

Medvedev and Melott (2007) noted that the Sun and planets not only orbit the galaxy but oscillate perpendicular to the galactic disc with a period in the vicinity of 60 Myr. Computation of the period is uncertain, owing to limited knowledge of the mass density of the galaxy, including dark matter. Medvedev and Melott hypothesized that radiation effects associated with a plasma shock front on the side of the galaxy toward the Virgo cluster might induce a periodic fluctuation in diversity. Melott et al. (2010) showed that resulting atmospheric ionization, giving rise to increased ultraviolet radiation on the earth, might be sufficient to stress the biosphere. Atri and Melott (2011), and Rodriguez et al. (2013) showed that muons produced by these cosmic rays reaching the surface may constitute a further radiation hazard and reduce photosynthesis. It is also possible that passing through the galactic disk every half-period might have a relationship to the 27-Myr extinction pulse discussed here, but no robust mechanism has been identified for this.

Melott and Bambach (2011b) and Meyers and Peters (2011) noted periodicities in the vicinity of 60 Myr in the emplacement of sedimentary packages. Melott et al. (2012) observed a strong ~60-Myr periodicity in $^{87}Sr/^{86}Sr$, which is strikingly coincident with biodiversity fluctuations.  Prokoph et al. (2013) and Rampino and Prokoph (2013) confirmed this and further noted a coincidence with the emplacement of large igneous provinces, all suggesting a possible tectonic effect driving the longer period fluctuation.  However, none of these show spectral power near a 27-Myr period—although some of them would be hard pressed to do so given the limitations of time resolution.

The combination of two periodic systems, each capable of promoting increased extinction, shifting in and out of phase and coupled to the variability in severity of multiple types of ecologically disruptive events playing out as the global fauna changed radically in dominant taxa over time (Wang and Bush 2008; Bush and Bambach 2011), makes the comparative study of extinction difficult and the determination of the causal nexus responsible for any individual extinction event difficult as well. This complexity is the major reason why definitive conclusions from the comparative study of extinction events are so elusive.

### Conclusions

1.  Time series analysis of extinction intensity recorded independently in the Paleobiology Database and the Sepkoski compendium provide significant evidence of a periodicity in extinction close to 27 Myr extending over at least the last 470 Myr. This periodicity may continue further back in time, but correlation issues and radical differences in evolutionary dynamics in the Cambrian prevent direct analysis.

2.  A feature at 27 Myr that had been noted in the spectrum of interval lengths using the 2004 timescale (e.g., Lieberman and Melott 2007) is no longer present when dates from the 2012 timescale are used. However, the continued presence of the periodicity in extinction in two different data sets when improvements in the timescale eliminate the slight evidence of a 27-Myr feature in the 2004 timescale is a strong argument for the reality of the periodicity.

3.  The timing of the 19 intervals of marked extinction intensity which occur in the last 470 Myr include a significant excess of intervals within ±3 Myr of the 27-Myr timing of peak extinction; and these events significantly avoid the "gap"—at the trough in extinction halfway between the 27-Myr peaks of extinction. These intervals of marked extinction intensity also strongly prefer the decreasing diversity phases of a previously identified 62-Myr periodicity in diversity.

4.  These two periodic systems appear to increase severity of extinction by modifying the effects of the immediate disruptive events commonly thought of as the causes of extinction.

5.  Because the effects of environmental disruption are dependent on the waxing and waning of the two periodicities as well as the shifting sensitivity to extinction of the global fauna produced by evolutionary turnover in dominant taxa, previous lack of understanding of these periodicities may be a principal reason why it has been difficult to do comparative study of extinctions. Full understanding of the interactive effects of all three phenomena (rates of environmental disruption, change in faunal composition, and timing causes of periodic influences) will be needed to adequately understand the role of extinction mechanisms in the history of life.

## Acknowledgments

We thank J. Alroy for providing his 2008 extinction data, and M. Foote, B. Lieberman, M. Patzkowsky, D. Raup, and an anonymous referee for comments that helped improve the presentation and the clarity of the paper. This is Paleobiology Database publication number 189.

TABLE 1. Data on intervals of marked extinction intensity.

| Period | Event or substage | Age (Ma) | Proportion of genus extinction | Proportional diversity change | 62-Myr phase | Myr from 27-Myr spacing | | |
|--------|-------------------|----------|-------------------------------|-------------------------------|--------------|-----|---------|-----|
| | | | | | | Hit | Neither | Gap |
| **The "Big Five" mass extinctions** | | | | | | | | |
| Cretaceous | End-Maastrichtian | 66.00 | 0.404 | -0.332 | Decrease | 2.18 | | |
| Triassic | Rhaetian event | 201.30 | 0.431 | -0.266 | Decrease | 1.68 | | |
| Permian | Changhsingian event | 252.20 | 0.557 | -0.463 | Decrease | 1.74 | | |
| Devonian | Late Frasnian event | 372.20 | 0.347 | -0.213 | Decrease | | 9.62 | |
| Ordovician | Late Hirnantian event | 443.80 | 0.305 | -0.251 | Decrease | 0.26 | | |
| Ordovician | Katian/Hirnantian event | 445.20 | 0.397 | -0.226 | Decrease | 1.14 | | |
| **Other mass extinctions (Bambach 2006)** | | | | | | | | |
| Neogene | Pliocene events | 2.59 | 0.084 | 0.044 | Decrease | | 6.91 | |
| Paleogene | Late Eocene | 33.90 | 0.156 | -0.011 | Increase | 2.76 | | |
| Cretaceous | Late Cenomanian event | 93.90 | 0.136 | -0.008 | Increase | 2.92 | | |
| Jurassic | Late Tithonian | 145.00 | 0.199 | -0.060 | Decrease | 0.30 | | |
| Jurassic | Early Toarcian event | 182.00 | 0.176 | -0.010 | Decrease | | 9.54 | |
| Permian | Capitanian event | 259.80 | 0.481 | -0.317 | Decrease | | 5.86 | |
| Carboniferous | Mid-Serpukhovian event | 327.05 | 0.309 | -0.132 | Decrease | | 8.37 | |
| Devonian | Late Famennian events | 358.90 | 0.311 | -0.071 | Increase | | 3.68 | |
| Devonian | Late Givetian events | 382.70 | 0.285 | -0.107 | Decrease | | 7.04 | |
| **Intervals of no mass extinctions** | | | | | | | | |
| Neogene | Middle Miocene | 11.63 | 0.089 | 0.037 | Decrease | 2.13 | | |
| Triassic | Spathian | 247.10 | 0.389 | 0.141 | Decrease | | 6.84 | |
| Carboniferous | Late Moscovian | 307.00 | 0.203 | -0.001 | Decrease | 1.26 | | |
| Ordovician | Mid-Caradocian | 452.14 | 0.228 | 0.030 | Decrease | | 8.08 | |
| **Cambrian extinction events qualifying as intervals of marked extinction intensity** | | | | | | | | |
| Cambrian | End-Sunwaptan event | 486.00 | 0.525 | -0.212 | Increase | | | 12.38 |
| Cambrian | End-Steptoan event | 493.00 | 0.443 | 0.003 | Decrease | | 5.38 | |
| Cambrian | End-Marjuman event | 497.00 | 0.607 | -0.295 | Decrease | 1.36 | | |
| Cambrian | End-Toyonian event | 510.00 | 0.285 | -0.080 | Decrease | | | 11.62 |
| Cambrian | End-Botomian event | 513.00 | 0.540 | -0.305 | Decrease | | | 12.54 |
| Cambrian | End-Tommotian event | 521.00 | 0.591 | 0.193 | Decrease | | 4.54 | |

**Figures**

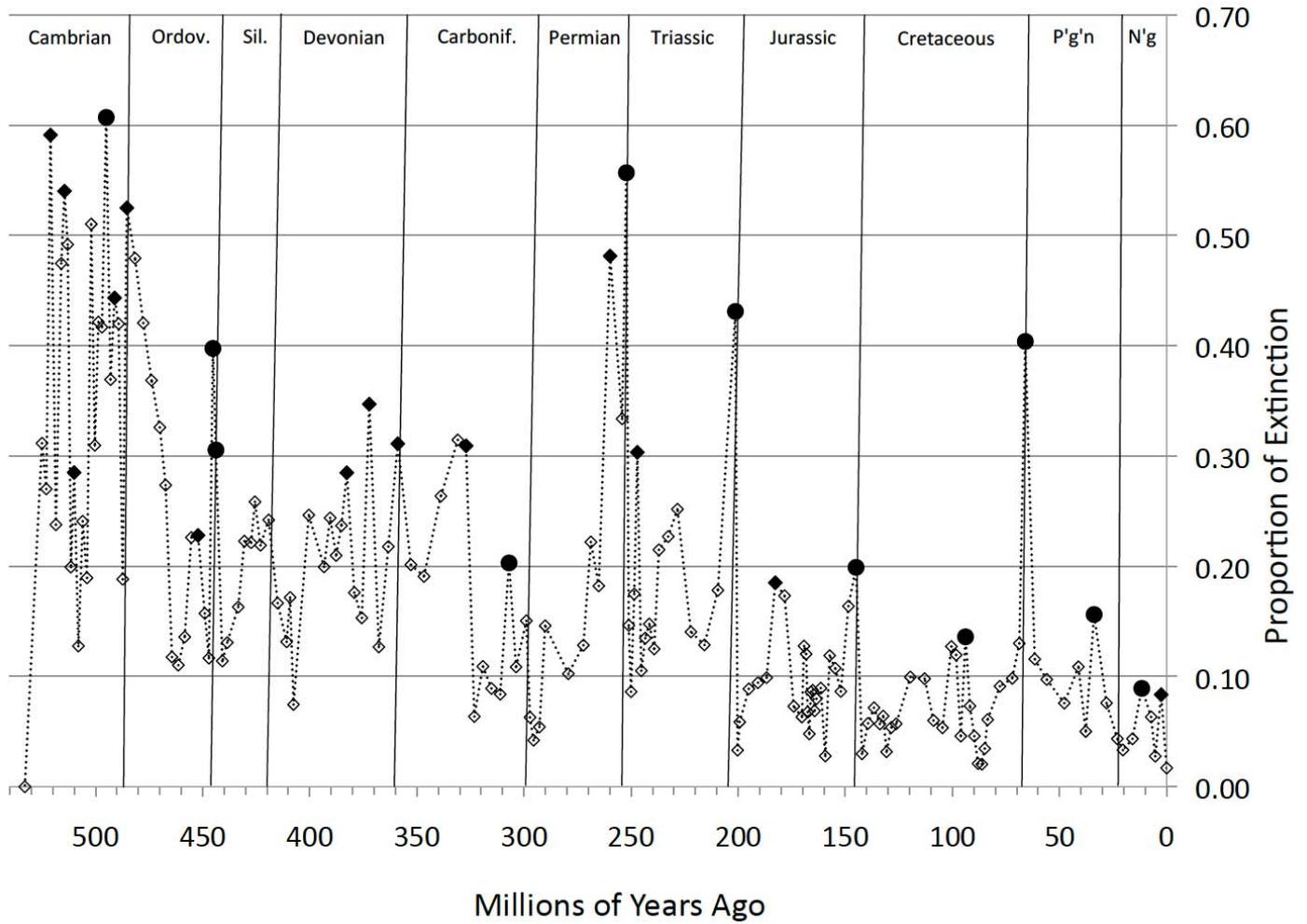

FIGURE 1. Fractional (proportional) extinction rate by substage for Sepkoski genus data. Filled symbols are the 25 intervals of marked extinction intensity noted in the text. The larger filled circles mark the intervals of marked extinction intensity that lie within ±3 Myr of the maxima of the 27-Myr periodicity as documented in the text

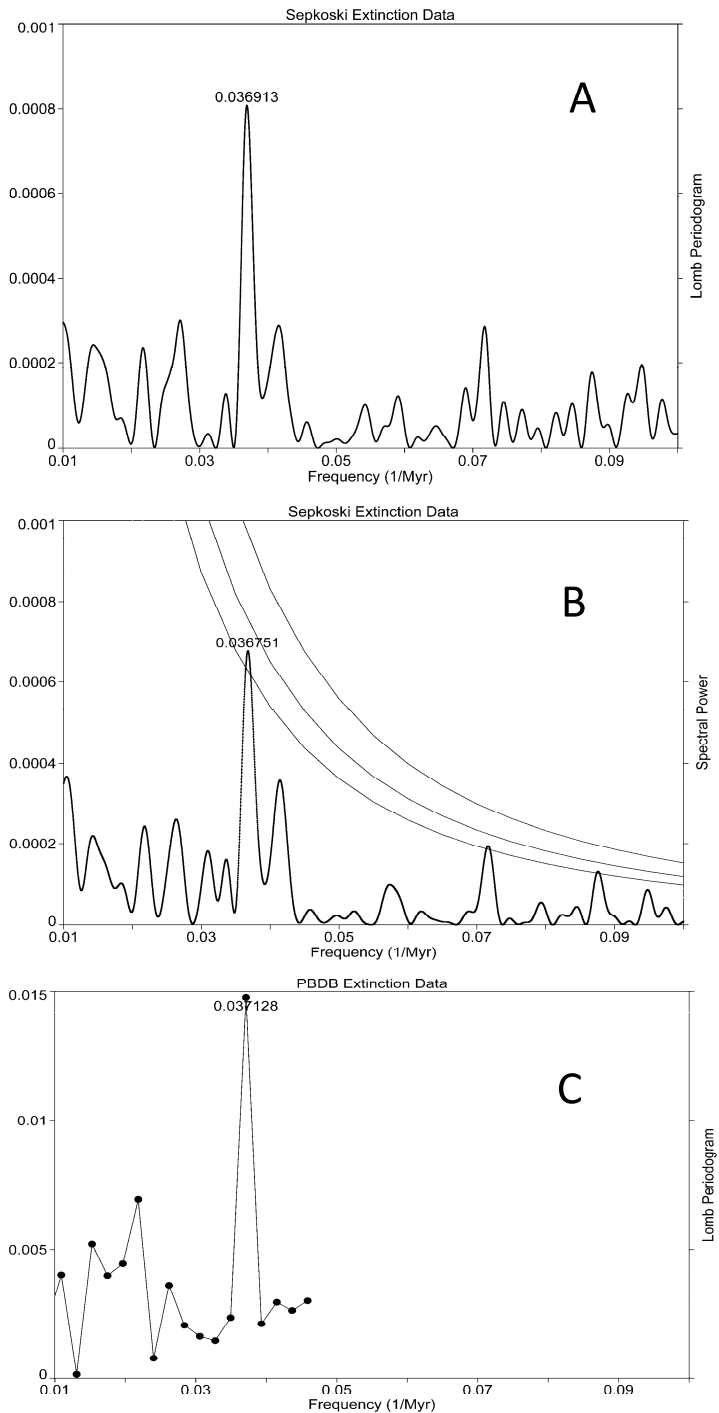

FIGURE 2. A, Lomb periodogram of the Sepkoski data. B, Power spectrum of extinction in the Sepkoski data, computed by FFT. The noted peak is at about 27.2 Myr. The curved lines indicate confidence levels of .05, .01, and .001. C, Lomb periodogram of the Paleobiology Database extinction rate (see text). There is a spectral peak at 27.0 Myr. The difference in amplitude between the two plots is not significant, as it arises from a different procedure for defining extinction rates as given in the two data sets. The 2012 geological timescale is used and extinction data are assigned dates at the interval end boundaries.

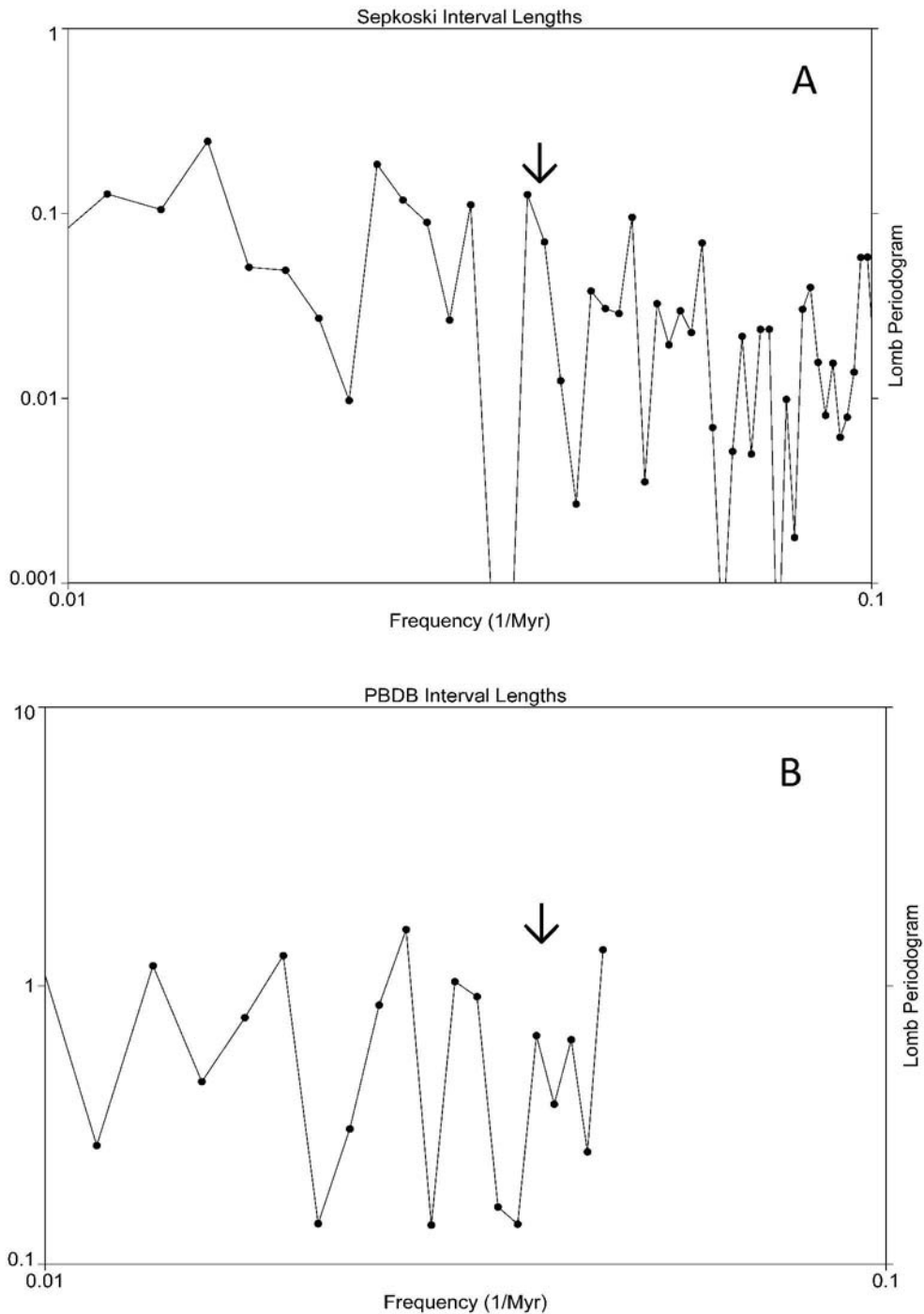

FIGURE 3. A, Lomb periodogram of the interval lengths for the substages used in the Sepkoski data using the 2012 geological timescale. A spectral peak at the frequency corresponding to 27 Myr (marked by downward arrow) exists but is no greater than any others. B, Lomb periodogram of the interval lengths for the binning intervals of the PBDB. No peak occurs at 27 Myr—which is again marked by a downward arrow.

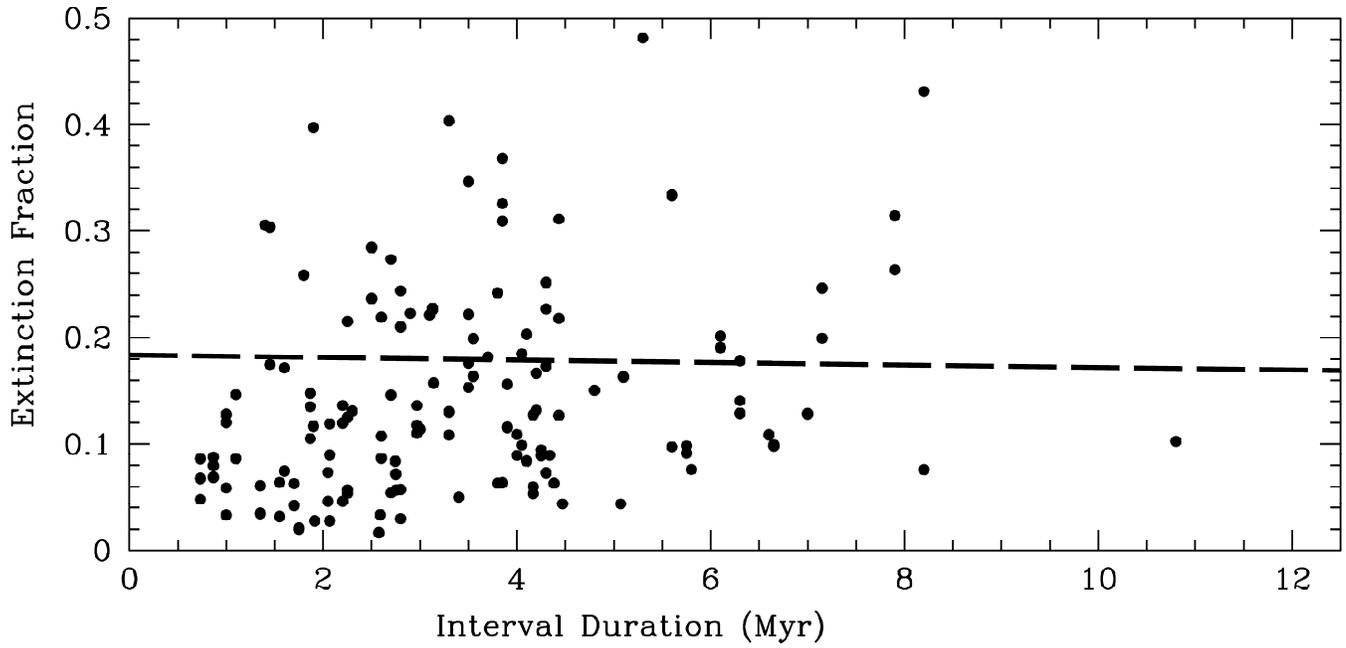

FIGURE 4. We plot the fraction of extinction in the SEP data against the length of the interval over the last 475 Myr. The dashed line is a least squares linear fit $0.184 - 0.0012D$, where $D$ is the interval length in Myr. This supports the contention that most extinction is pulsed rather than continuous, although highly variable as indicated by the large scatter.

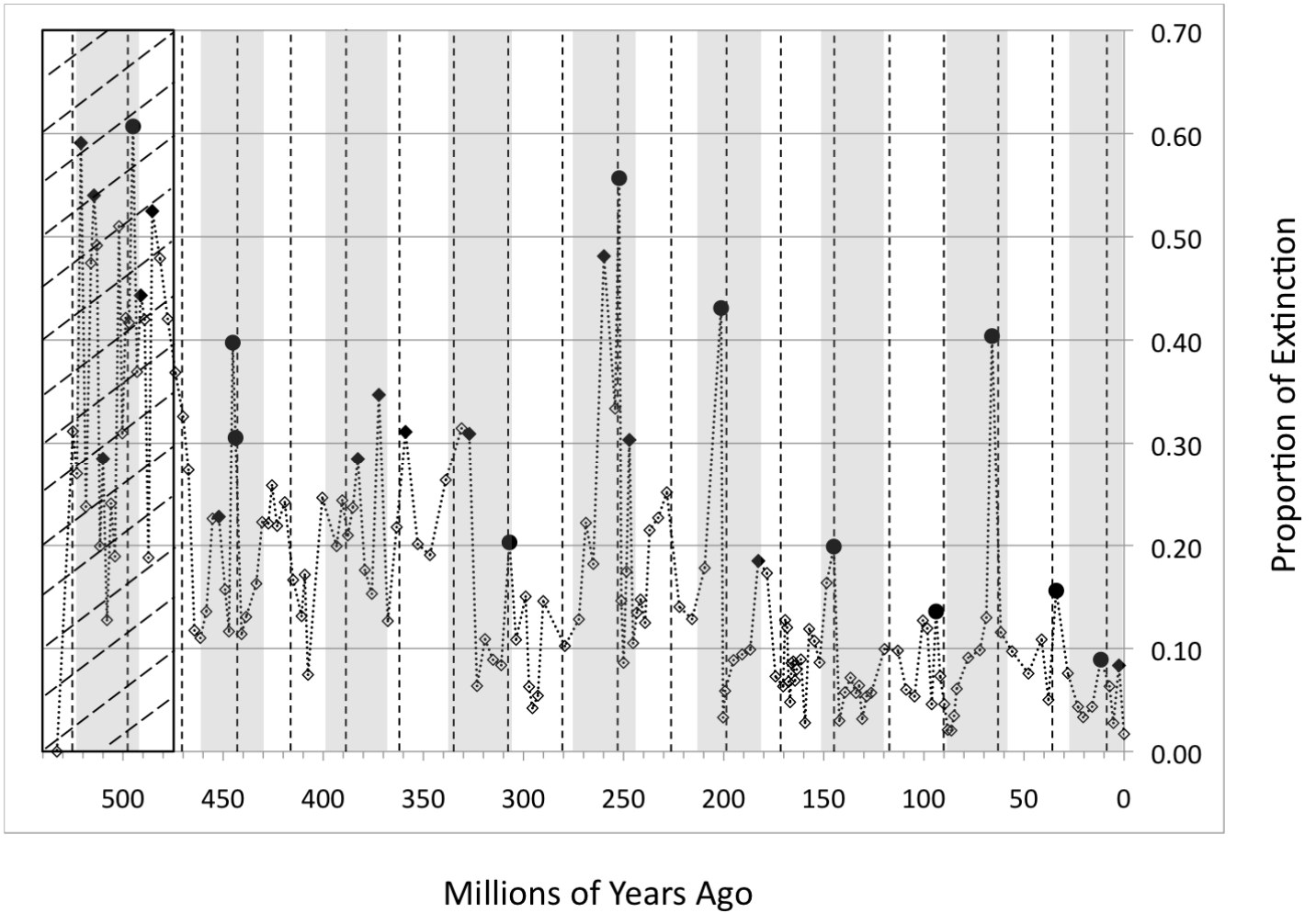

FIGURE 5. Fractional (proportional) extinction rate for the Sepkoski data. Shaded vertical bands are the 31-Myr decreasing diversity phases of the 62-Myr periodicity and unshaded vertical bands are the increasing diversity phases. Dashed vertical lines are the maxima in the 27-Myr periodicity of extinction. Filled symbols are the 25 intervals of marked extinction intensity. The larger filled circles mark the intervals of marked extinction intensity that lie within ±3 Myr of the maxima of the 27-Myr periodicity. The area highlighted by slanting dashed lines at the left is the portion of the Phanerozoic characterized by unusually elevated evolutionary dynamics and was not used to determine the periodicities. The six filled symbols in the Cambrian are for extinction events noted in Gradstein et al. (2012) that qualify as intervals of marked extinction intensity.